\documentstyle[twocolumn,aps,epsfig]{revtex}

\newcommand{\be}{\begin{eqnarray}}
\newcommand{\ee}{\end{eqnarray}}

\begin{document}
\draft

\title{
Large-$p_T$ Inclusive $\pi^0$ Production in Heavy-Ion 
Collisions at RHIC and LHC
}
\author{Sangyong Jeon$^{1,2}$, Jamal Jalilian-Marian$^3$ and Ina Sarcevic$^4$}
\address{
$^1$RIKEN-BNL Research Center, Upton, NY 11973-5000\\
$^2$Department of Physics, McGill  University, Montreal, 
QC H3A-2T8, Canada \\
$^3$Physics Department, Brookhaven National Laboratory, Upton, NY 11973, USA 
11794\\
$^4$Department of Physics, University of Arizona, Tucson, Arizona
85721, USA\\
}

\wideabs{

\maketitle

\begin{abstract}

\widetext 

We present results for the large-$p_T$ inclusive $\pi^0$ production in 
p-p and A-A collisions  at RHIC and LHC energies.  We include the full 
next-to-leading order radiative corrections, $O(\alpha_s^3)$, and nuclear 
effects such as parton energy loss and nuclear shadowing. We find the 
next-to-leading order corrections and the parton energy loss effect to be 
large and $p_T$ dependent, while the nuclear shadowing effects are small 
($< 10\%$). We calculate the ratio of prompt photons to neutral pions 
produced in heavy ion collisions and show that at RHIC energies this ratio 
increases with $p_T$ approaching one at $p_T \sim 8$ GeV, due to the large 
suppression of $\pi^0$ production. We show that at the LHC, this ratio has 
a steep $p_T$ dependence and approaches $25\%$ effect at $p_T \sim 40$ GeV.  
We discuss theoretical uncertainties inherent in our calculation, such as 
choice of the renormalization, factorization and fragmentation scales and 
the K-factors which signify the size of higher-order corrections.  

\end{abstract}
}

\vskip 0.1true in

\narrowtext
\section{Introduction}

In high energy heavy ion collisions hard scatterings of partons occur 
in the early stages of the reaction, well before the possible formation of
a quark gluon plasma, resulting in production of large 
transverse momentum particles.  Fast partons produced in these 
hard collisions propagate through the hot and dense medium created in the
heavy ion collision and lose their energy. Therefore, a possible 
signature for parton energy loss in a hot, dense medium is the suppression 
of pion production in heavy ion collisions relative to 
hadron-hadron collisions. The Relativistic Heavy-Ion Collider (RHIC) at 
BNL with Au-Au collisions at $\sqrt{s}=130$ GeV and at $\sqrt{s}=200$ GeV, 
and the Large Hadron Collider (LHC) at CERN which will collide Pb-Pb at 
$\sqrt{s}=5.5$TeV provide the best opportunity to study the properties of 
the hot and dense matter and the possible formation of a new phase, the 
quark-gluon plasma. Recent measurements of inclusive $\pi^0$ 
production at RHIC energy of $\sqrt{s}=130$ GeV \cite{rhic} and at 
$\sqrt{s}=200$ GeV \cite{rhic200} show a large suppression in the $p_T$ 
region ($1$GeV$ < p_T < 10$GeV) which has created much excitement in the 
field.  The question of the origin of this suppression has inspired many 
interpretations. Clearly, the low $p_T$ region has large nonperturbative 
contributions and perturbative calculation cannot be trusted. Here we 
concentrate on production of $\pi^0$ in the large-$p_T$ region where 
perturbative QCD calculations are expected to be more reliable. 

In addition to being of interest for studying nuclear effects, such as 
parton energy loss and nuclear shadowing, large-$p_T$ $\pi^0$ mesons 
form a significant background for the prompt photons.  In principle, 
large-$p_T$ pions could form ``fake prompt photons'' when one photon 
from pion decay escapes detection. Theoretical predictions 
for the ratio of prompt protons to pions at RHIC and LHC energies are 
crucial for studying possible quark-gluon plasma formation via photons. 
In addition, this ratio may reduce some of the theoretical uncertainties, 
such as choice of factorization, renormalization and fragmentation scale, 
or the choice of gluon fragmentation function.  

In perturbative QCD, the inclusive cross section
for pion production in a hadronic collision is given by: 

\be
E_\pi \frac{d^3\sigma}{d^3p_\pi}(\sqrt{s},p_\pi)
&=&
\int dx_{a}\int dx_{b} \int dz \sum_{i,j}F_{i}(x_{a},Q^{2}) \nonumber \\
&&
F_{j}(x_{b},Q^{2}) D_{c/\pi}(z,Q^2_f) E_\pi 
{d^3\hat{\sigma}_{ij\rightarrow c X}\over d^3p_\pi} 
\label{eq:factcs}
\ee

\noindent
where $F_{i}(x,Q^{2})$ is the i-th parton distribution in a nucleon,
$x_a$ and $x_b$ are the fractional momenta of incoming partons,
$D_{c/\pi}(z,Q^2_f)$ is the pion fragmentation function, $z$ is the fraction 
of parton energy carried by the pion and 
${d^3\hat{\sigma}_{ij\rightarrow c X}\over d^3p_c}$ are parton-parton cross 
sections which include leading-order, $O(\alpha_s^2)$, subprocesses such as: 
\be
q + q &\rightarrow &  q + q  \nonumber\\
q + \bar q &\rightarrow & \bar q + q  \nonumber\\
q +  g &\rightarrow & g   + q \nonumber\\
g + g &\rightarrow & g  + g   
\ee

\noindent 
and the next-to-leading order, $O(\alpha_s^3)$, subprocesses such as:
\be
q + q &\rightarrow &  q + q + g  \nonumber\\
q + \bar q &\rightarrow & q + \bar q + g  \nonumber\\
q + q' &\rightarrow & q + q' + g  \nonumber \\
q + \bar q &\rightarrow & q' + \bar q' +  g  \nonumber\\
g + g &\rightarrow & g + g + g 
\ee

The running coupling  constant $\alpha_{s}(\mu^{2})$, calculated to
next-to-leading order, is given by

{
\small
\begin{eqnarray*}
\alpha_s (\mu^2)={12\pi\over (33-2N_f)\ln \mu^2/\Lambda^2}
\bigg[1-{6 (153-19N_f)\ln\ln \mu^2/\Lambda^2 \over
(33-2N_f)^2\ln \mu^2/\Lambda^2}\bigg] \nonumber\\
\end{eqnarray*}
}

\noindent
where $\mu$ is the renormalization scale, $N_F$ is the
number of flavors and $\Lambda$ is the $\Lambda_{QCD}$ scale.

The parton distribution functions $F_{i}(x,Q^{2})$ are measured in Deep
Inelastic Scattering experiments such as those at HERA \cite{hera} while
fragmentation functions, $D_{c/\pi}(z,Q^2_f)$, that describe the transition 
of the partons into the final-state pions are extracted from $e^+e^-$ 
annihilation data from PETRA, PEP and LEP \cite{frag}.  
The gluon fragmentation function, which gives the dominant contribution 
to $\pi^0$ production at the LHC, is not well determined by $e^+e^-$ data, 
since it appears only at NLO.
Nevertheless, it is possible to get some 
constraint on the gluon fragmentation function from measurements of large 
$p_T$ pion production in hadronic collisions at high energies. Such a study 
has been done using UA1 data in the range $5$GeV$ <p_T<20 $GeV \cite{UA1}.  
The gluon fragmentation function of 
Binnewies, Kniehl and Kramer (BKK) \cite{frag} 
 is found to be consistent with 
the UA1 data.  
The fragmentation scale, $Q_f$, when taken to be too small, i.e. 
$Q_f=p_T/3$, probes the region currently not tested by the data. In 
addition, theoretical improvement is needed in resumming large $ln(1-z)$ 
terms present in the higher-order corrections.  

Divergencies present 
in the calculation of parton cross section,
$d\hat{\sigma}_{ij}$, require careful
treatment of collinear and ``soft'' singularities in the
matrix elements.
Divergences cancel between real and virtual graphs when the
physical $\pi^0$ cross section is calculated.
Infrared or ``soft'' divergences
($k\rightarrow 0$) cancel between virtual amplitudes
and real amplitudes.
Collinear divergences ($\theta \rightarrow 0$ between initial quark and
radiated gluon, for example) are reabsorbed
in structure function and fragmentation function 
renormalization (``mass'' singularities).

Detailed study of large-$p_T$ inclusive $\pi^0$ production in 
hadronic collisions show very good agreement between theory and experiments, 
apart from an overall normalization \cite{se}.  

\section{Large-$p_T$ Inclusive $\pi^0$ Production in Heavy Ion Collisions} 

To calculate the inclusive cross section for pion production in heavy
ion collisions, we will use (\ref{eq:factcs}) with the distribution
and fragmentation functions appropriately modified to include nuclear
effects such as shadowing and energy loss.

\begin{figure}[htp]
\centering
\setlength{\epsfxsize=7.5cm}
\centerline{\epsffile{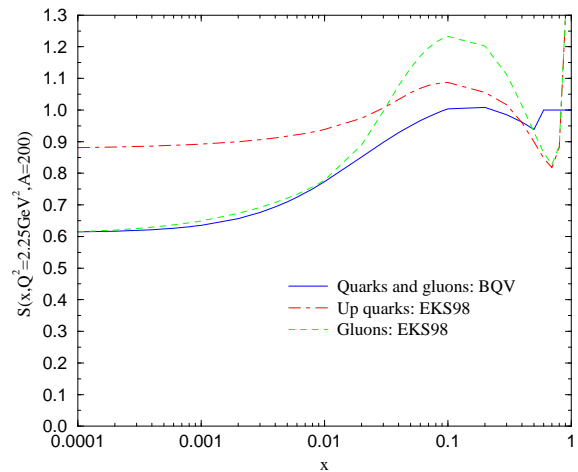}}
\caption{The nuclear shadowing ratio as parametrized by BQV 
and EKS98.}
\label{fig:eksbqv}
\end{figure}

It is a well known experimental fact that the distribution of quarks 
and gluons inside nuclei is modified compared to that in a free nucleon. 
This modification is known as nuclear shadowing (for a review of nuclear 
shadowing, we refer the reader to \cite{arneodo}). The parton 
distribution in a nucleus $F_{a/A}(x,Q^2,b_t)$, can be written as 
\begin{eqnarray*}
F_{a/A}(x,Q^2,b_t)=T_A(b_t)\,S_{a/A}(x,Q^2)\,F_{a/N}(x,Q^2)
\end{eqnarray*}
\noindent
where $T_A(b_t)$ is the nuclear thickness function, $F_{a/N}(x,Q^2)$ is the
parton distribution function in a nucleon and $S_{a/A}(x,Q^2)$ is the
parton shadowing function, 
$S_{a/A}(x,Q^2) = F_{a/A}(x,Q^2)/A F_{a/N}(x,Q^2)$.  
  It should be emphasized that the $Q^2$ dependence
of nuclear shadowing is poorly known, especially in the small $x_{bj}$ 
region. Also, gluon shadowing is measured only indirectly through scaling
violation of $F_2$ structure function which leads to large uncertainties
in the gluon distribution function in nuclei. In this work, we use 
the nuclear shadowing parametrization due to Eskola, Kolhinen and 
Salgado (EKS98),
which is $Q^2$ dependent and distinguishes between quarks and gluons
\cite{eks98} and was shown to be in very good agreement with the NMC data
on $Q^2$ dependence of $F_2^{Sn}/F_2^C$ \cite{NMC}, while some
other parametrizations which have large gluon suppression due to
nuclear shadowing are ruled out \cite{ehks}. It is also shown that
modifications to the DGLAP evolution due to gluon fusion are small for
the kinematic region of relevance to RHIC. In Fig. (\ref{fig:eksbqv}) we 
show EKS98 nuclear shadowing function at $Q^2=2.25 $GeV$^2$ and, for 
comparison, we also show nuclear shadowing parametrization 
due to Benesh, Qiu and Vary (BQV) \cite{bqv} which is $Q^2$ independent 
and treats shadowing of quarks and gluons on the same footing. 

\begin{figure}[htp]
\centering
\setlength{\epsfxsize=7.5cm}
\centerline{\epsffile{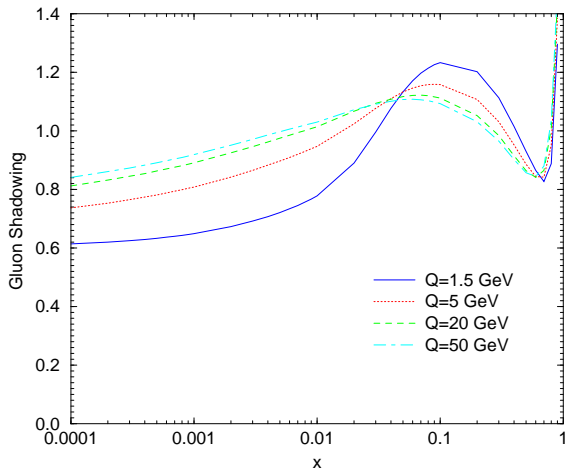}}
\caption{$Q^2$ dependence of gluon shadowing function 
in a nucleus ($A=208$) in EKS parametrization.}
\label{fig:shadQ2}
\end{figure}

In Fig. (\ref{fig:shadQ2}) we show the $Q^2$ dependence of the
shadowing function for gluons in EKS parametrization
\cite{ekrs}.  Clearly, there is a strong $Q^2$ dependence
in the EKS parametrization of nuclear shadowing, especially
for gluons, as well as large anti-shadowing effect 
in the large $x_{bj}$
region.

Another nuclear effect which we include is the medium induced 
energy loss. It is expected that fast partons propagating through
the hot and dense medium created after a high energy heavy ion
collision will scatter from the partons in the medium, lose
part of their energy and then fragment into hadrons with a reduced 
energy. As a result, the spectrum of the final state hadrons observed
in a heavy ion collision is expected to be suppressed compared to
hadronic collisions. This has been extensively studied and we refer
the reader to \cite{eloss} for some recent work. The 
Landau-Pomeranchuk-Migdal (LPM) effect occurs when an ultrarelativistic 
particle emits low energy bremsstrahlung photons (gluons) as the particle 
passes through dense matter; fewer photons (gluons) are emitted than 
predicted by bremsstrahlung theory for isolated atoms.  

While a dynamical study of the parton propagation in a hot and dense medium
created in a realistic heavy ion collision and the modification of the 
hadronization is more desirable, we will use a phenomenological model 
\cite{hsw} here to demonstrate how sensitive pion production is to the 
average energy loss suffered by a parton in a hot and dense medium.
We restrict ourselves to the central rapidity region so that a parton will 
only propagate in the transverse direction in a cylindrical system. Given the 
inelastic scattering mean-free-path, $\lambda_a$, the probability for a 
parton to scatter $n$ times within a distance $\Delta L$ before it escapes 
the system is assumed to be given by the Poisson distribution

\be
P_a(n) = \frac{(\Delta L/\lambda_a)^n}{n!} e^{-\Delta L/\lambda_a}.
\nonumber 
\end{eqnarray}

In the Huang-Sarcevic-Wang model of energy loss, the hadronic 
fragmentation function $D_{c/\pi}(z,Q^2)$ is modified in order to include 
multiple scattering of a parton in a nuclear medium. If the average energy 
loss per scattering suffered by the parton $a$ is $\epsilon_a$, the nuclear
fragmentation functions can be modeled as \cite{hsw}, 

\be
z D_{c/\pi}(z,\Delta L,Q^2)& =&
 \sum_{n=0}^NP_a(n) z^a_n D^0_{c/\pi}(z^a_n,Q^2)
  \nonumber \\
&+&\langle n_a\rangle z'_aD^0_{g/\pi}(z'_a,Q_0^2), 
\label{eq:mfrag}
\ee

where $z^a_n=z/(1-n\epsilon_a/E_T)$, $z'_a=zE_T/\epsilon_a$, $N$ is 
the maximum number of collisions for which $z_n^a \le 1$ and 
 $D^0_{c/\pi}$ is the hadronic 
fragmentation function which gives the probability that quark or a gluon 
would fragment into a pion. 
The first term corresponds to the fragmentation of the leading partons 
with reduced energy $E_T-n\epsilon_a$ and the second term comes from the 
emitted gluons each having energy $\epsilon_a$ on the average. 
The average number of scatterings within a distance $\Delta L$
is $\langle n_a\rangle =  \Delta L/\lambda_a$.  We take 
$\lambda_a=1fm$ and $\Delta L= R_A$.  

In this work, we use the NLO code due to Aurenche {\it et al.} for 
calculating $\pi^0$ and prompt photon production in p-p collisions 
\cite{se} with MRS99 parametrization of nucleon structure functions 
\cite{mrs99}and BKK pion fragmentation functions \cite{frag}.  We use 
 EKS98 
shadowing functions to include nuclear shadowing and we modify the BKK 
fragmentation functions according to the model of Huang-Sarcevic-Wang 
\cite{hsw} in order to take into account medium induced parton\cite{se}  
energy loss effects. We calculate the invariant cross section for 
$\pi^0$ and prompt photon production in heavy-ion collision normalized 
to the number of binary nucleon-nucleon collisions, $N_{coll}$, where 
$N_{coll}$ can theoretically be determined from nuclear overlapping function, 
i.e.  $N_{coll}= T_{AA}(b)\sigma_{inel}^{NN}$ and $T_{AA}(b)= 
 \int d^{2}b_{1}
T_{A}(\mid\vec{b}_{1}\mid)
T_{A}(\mid\vec{b} - \vec{b}_{1}\mid)$.  
The number of N-N collisions depends on the centrality that experiment 
triggers on.  Here we take 
$N_{coll}=975$, which is obtained 
by PHENIX for their central collisions \cite{rhic200}.   
In order to investigate the sensitivity of our results to the 
choice of energy loss parameters in the model of \cite{hsw}, we consider 
various forms of the average parton energy loss parameter, $\epsilon$; 
constant energy loss as well as two different energy dependent forms of
energy loss.  We find that fractional energy loss, 
${\Delta E \over E} = 6.1\%$ gives the best description of the $\pi^0$ data
at RHIC for $\sqrt{s}=200 GeV$ (see Fig. 8).   With this energy loss, 
we make predictions for prompt photon invariant cross sections. We set all 
the scales, factorization, renormalization and
 fragmentation scale to be equal, 
$Q = Q_F = \mu = \kappa p_T$ where we take $\kappa=1$ and $\kappa=2$.    
We discuss sensitivity of our results to the choice of scale\footnote{We do 
not consider $\kappa=0.5$ because for low $p_T$ the scales would be in the 
region where currently there is no data and where 
 perturbative QCD in not reliable.} We assume that nucleus consists of $A$ 
protons, i.e. we do not take into account isospin effects, 
which are expected to be small for large $A$.
 
\begin{figure}[htp]
\centering
\setlength{\epsfxsize=7.5cm}
\centerline{\epsffile{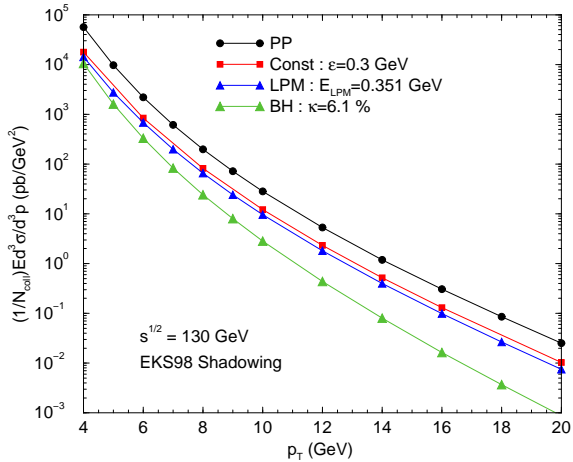}}
\caption{Inclusive $\pi^0$ distribution at $\sqrt{s}=130$ GeV.}
\label{fig:rhic130dsigma}
\end{figure}

In Fig. (\ref{fig:rhic130dsigma}) we show the invariant cross section 
${E\, d^3\sigma \over d^3\,p}$ for inclusive $\pi^0$ production 
in proton-proton and nucleus-nucleus ($A=200$) collisions at 
$\sqrt{s}=130$ GeV.   
There is a clear suppression of the nuclear cross
section compared to the hadronic one. While constant energy loss of 
$\epsilon =0.3 GeV$ and LPM type energy loss $\epsilon \sim \sqrt{E}$
show similar shape of $p_T$ spectrum, fractional energy loss (BH) leads to a
much steeper spectrum. We choose values of $\epsilon$ and $E_{LPM}$ such that 
the suppresion at $p_T=4$GeV is in agreement with the data.  
However, the data at $\sqrt{s}=130 GeV$ is not precise enough to 
distinguish between the different forms of energy loss.
Nuclear shadowing effect, which was incorporated using 
EKS98 shadowing,  is found to be very small (few percent) at RHIC energies.  

From Fig. (\ref{fig:rhic130dsigma}) we see that 
the $\pi^0$ production cross section in heavy
ion collisions is suppressed compared to hadronic collisions even at very
large $p_T$ ($p_T>10$ GeV). This is due to the fact that high $p_T$ partons 
lose their energy as they propagate through the dense medium and contribute 
to the lower $p_T$ pions. 

\begin{figure}[htp]
\centering
\setlength{\epsfxsize=7.5cm}
\centerline{\epsffile{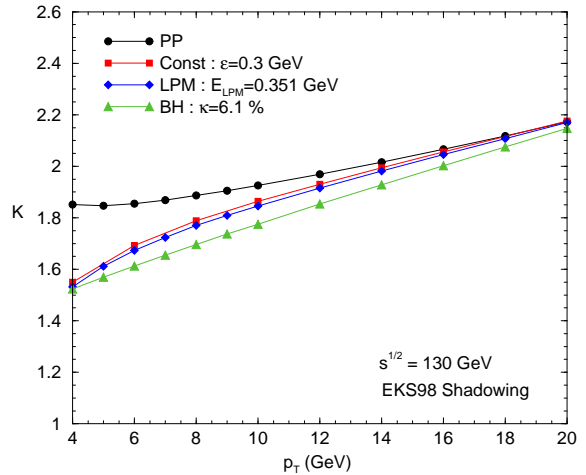}}
\caption{K-factor, defined as the ratio of full NLO and LO 
$\pi^0$ inclusive cross section at $\sqrt{s}=130$ GeV.}
\label{fig:rhic130k}
\end{figure}

The next-to-leading order contributions, $O(\alpha_s^3)$, which are included 
in our calculations, are large and $p_T$ dependent. For example, the nuclear 
K-factor defined as a ratio of full next-to-leading order calculation to the 
leading order is varying between 1.6 and 2.2 for $p_T$ between 4GeV and 20GeV, 
independent of the form of parton energy loss used.  
This is shown in Fig. (\ref{fig:rhic130k}). One should keep in mind that 
next-to-leading order calculation in nuclear collisions is not as complete 
as for the pp case due to the fact that EKS nuclear shadowing function is 
based on the leading-order DGLAP evolution and the parton energy loss has 
been calculated only in the leading-order approximation.  Uncertainties 
in the modified fragmentation function are most likely much larger than 
the uncertainty due to the 
next-to-leading order corrections to the nuclear shadowing function.

\begin{figure}[htp]
\centering
\setlength{\epsfxsize=7.5cm}
\centerline{\epsffile{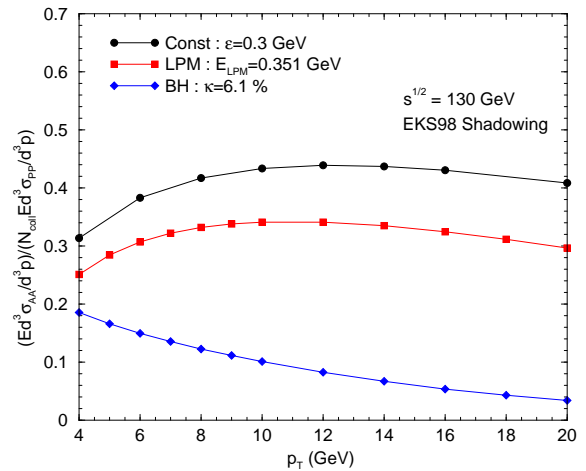}}
\caption{Ratio of inclusive $\pi^0$ distributions in 
heavy ion and p-p collisions at $\sqrt{s}=130$ GeV.}
\label{fig:rhic130ratio}
\end{figure}

In Fig. (\ref{fig:rhic130ratio}) we show the ratio of inclusive $\pi^0$ 
cross section in nucleus-nucleus collisions to the one in p-p collisions.  
Since the nuclear shadowing effects are small,  most of the
observed suppression is due to energy loss effects. The difference between
different forms of energy loss is striking, specially in the low to
intermediate $p_T$ region where they have different slopes. This difference
is even more pronounced for the case of fractional energy loss where this 
ratio actually decreases.  However, we point out that we do not
expect our rather crude model of energy loss through modified fragmentation
functions to be correct at very high $p_T$. This is due to the fact that
with a constant fractional energy loss as in the BH case, the ratio of 
$AA$ and $pp$ cross sections will never become one as one goes to higher
momenta unlike the LPM case where fractional energy loss goes away as 
$1/\sqrt{E}$. Although we cannot determine exact value of $p_T$ at which 
our model will break down from first principles, it is clear that it will 
happen at higher values of $p_T$ for higher energies.

We calculate prompt photon production in heavy ion collisions using 
Aurenche {\it et al.} program for prompt photon 
production in p-p 
collisions that includes all next-to-leading order contributions \cite{se}.  
In prompt photon production there are
two types of subprocesses that contribute, direct processes and
bremsstrahlung processes, the later one being convoluted with the
fragmentation functions, 
$D_{c/\gamma}(z,Q^2_f)$, that describe the transition of the partons into
the final-state $\gamma$.  The fragmentation functions without medium effects 
 are extracted from
$e^+e^-$ data \cite{frag}.  We use  
MRS99 parametrization of nucleon structure functions 
\cite{mrs99}, EKS98 nuclear shadowing functions and we modify fragmentation 
functions in the same way as for pions.

\begin{figure}[htp]
\centering
\setlength{\epsfxsize=7.5cm}
\centerline{\epsffile{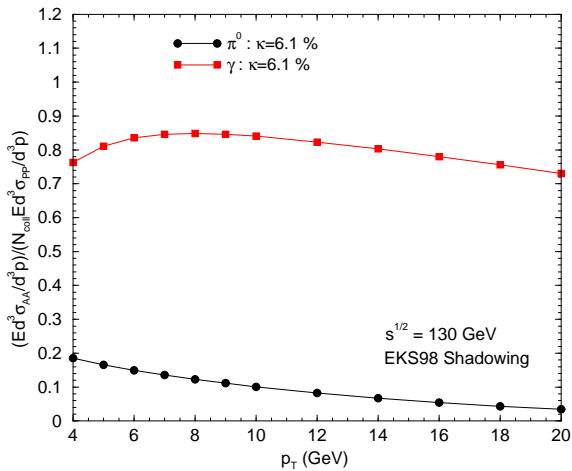}}
\caption{Prompt photon and inclusive $\pi^0$ cross sections in heavy ion 
collisions normalized to p-p at $\sqrt{s}=130$ GeV.}
\label{fig:rhic130gratio_pratio}
\end{figure}

In Fig. (\ref{fig:rhic130gratio_pratio}) we show the inclusive cross 
section for prompt photon production in heavy ion collisions at 
$\sqrt{s}=130$ GeV, normalized to proton-proton collisions, together with 
the same ratio for inclusive $\pi^0$ production. We note that 
for the same energy loss parameter $\epsilon = 6.1\% E$ , $\pi^0$ 
production is suppressed more than prompt photons. This is due to the
fact that at $p_T =3 GeV$, about $75\%$ of the photon contributions comes 
from ``direct'' processes which are not affected by parton energy loss, 
while in case of pions, all subprocesses are convoluted with the modified 
fragmentation function. At higher values of $p_T$, direct photons become
even more dominant at RHIC and their energy loss becomes negligible 
($< 10\%$ at $p_T=10 GeV$). Nuclear shadowing effect is small both in 
photon and in pion production.  
 
\begin{figure}[htp]
\centering
\setlength{\epsfxsize=7.5cm}
\centerline{\epsffile{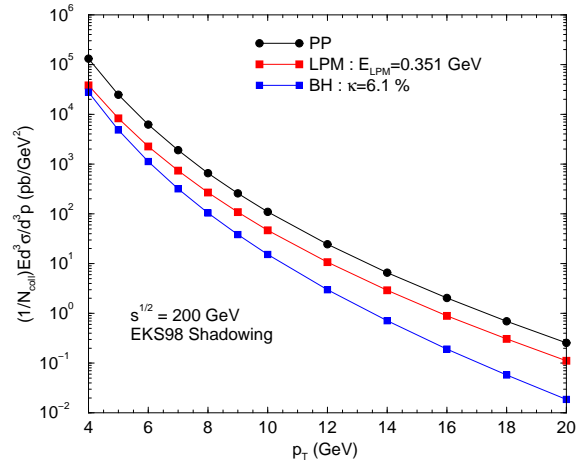}}
\caption{Inclusive $\pi^0$ spectrum at $\sqrt{s}=200$ GeV.}
\label{fig:rhic200dsigma}
\end{figure}

We show our results for inclusive $\pi^0$ production at RHIC at 
$\sqrt{s}=200$ in Fig. (\ref{fig:rhic200dsigma}) for different forms
of energy loss. Again, the difference between different energy loss
scenarios is significant especially at the highest $p_T$ where $\pi^0$ 
production differs 
by an
order of magnitude. We find that nuclear shadowing effects 
are small (less than $10\%$) and the suppression is mostly due to energy 
loss effects.

\begin{figure}[htp]
\centering
\setlength{\epsfxsize=7.5cm}
\centerline{\epsffile{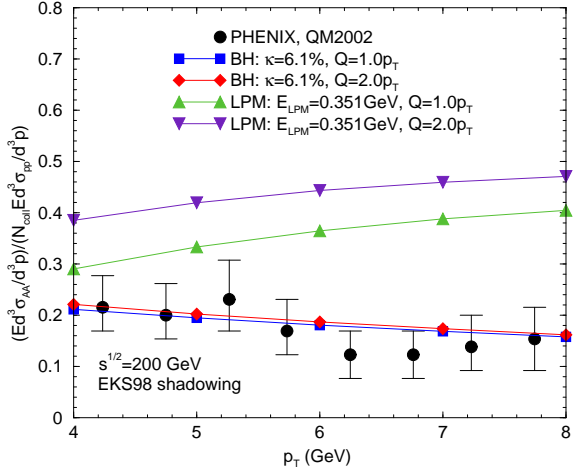}}
\caption{Ratio of inclusive $\pi^0$ cross sections in heavy ion and 
p-p collisions at $\sqrt{s}=200$ GeV, data is from PHENIX [2].}
\label{fig:rhic200ratio}
\end{figure}

In Fig. (\ref{fig:rhic200ratio}), we show the ratio of invariant cross
sections for $AA$ and $pp$ collisions at $\sqrt{s}=200 GeV$ for different
forms of energy loss. The $p_T$ spectrum behaves very differently in
the case of fractional energy loss as compared to the constant and LPM
type energy losses. Comparison to the preliminary PHENIX data \cite{rhic200} 
 seems to favor the fractional energy loss scenario.  Furthermore, 
varying the scales from $p_T$ to $2p_T$, in case of LPM energy loss, gives 
large uncertainty, about $30\%$, while in case of BH, this uncertainty is 
much smaller. This can be seen in Fig (\ref{fig:rhic200ratio}).  It is 
also interesting to note that change of scales does not affect the shape 
of the ratio.  

\begin{figure}[htp]
\centering
\setlength{\epsfxsize=7.5cm}
\centerline{\epsffile{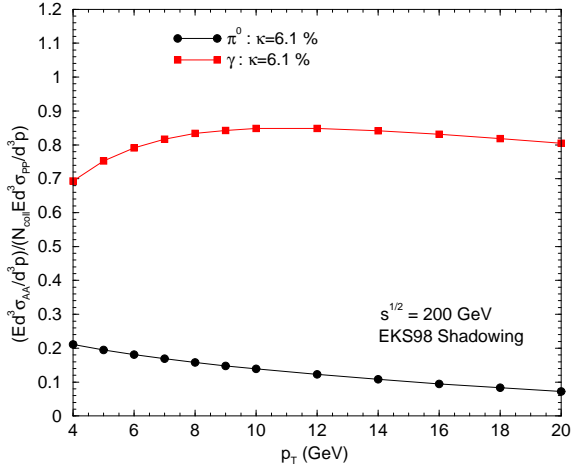}}
\caption{Prompt photon and inclusive $\pi^0$ cross sections in heavy ion 
collision normalized to p-p at $\sqrt{s}=200$ GeV.}
\label{fig:rhic200gratio_pratio}
\end{figure}

Prompt photon and pion invariant cross sections normalized to $pp$ collisions
are shown together in Fig. (\ref{fig:rhic200gratio_pratio}). Clearly, 
prompt photons are affected much less than pions due to direct photons
dominating over bremsstrahlung photons at high $p_T$. This is contrary to
expectations from approaches where the suppresion of high $p_T$ spectra
is due to high gluon density in the initial state (color glass condensate).
Therefore, measuring prompt photons would enable one to tell whether the
observed suppresion is due to initial state or final state effects.   
 
Measurements of inclusive pion production at RHIC energies for $p_T>4$ GeV 
could therefore provide valuable information about the medium induced 
parton energy loss since nuclear shadowing effects are very small 
(few $\%$) and most of the observed suppression of hadronic spectra
in heavy ion collisions is due to energy loss. 

\begin{figure}[htp]
\centering
\setlength{\epsfxsize=7.5cm}
\centerline{\epsffile{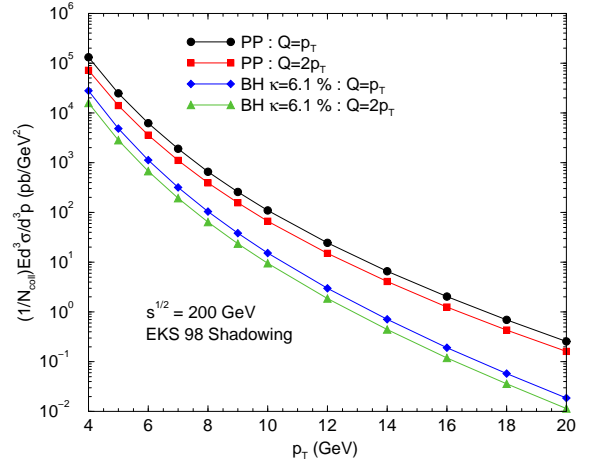}}
\caption{Theoretical uncertainty in inclusive $\pi^0$ cross section 
 due to choice of scale for $Q=p_T$ and 
$Q=2 p_T$. We assume that $Q=Q_F=\mu$.}  
\label{fig:uncerscale}
\end{figure}

In Fig. (\ref{fig:uncerscale}) we show theoretical uncertainty in 
predicting inclusive $\pi^0$ cross section by varying scales from 
$Q=p_T$ to $Q=2 p_T$, with the assumption that $Q=Q_F=\mu$.  

We note that uncertainty is about $40\%$ and that, most importantly, the 
shape of the $p_T$ spectrum is not very sensitive to the choice of 
scale (the uncertainty due to scale choices of $Q=2p_T$ and $Q=p_T/2$ are 
about $100\%$ even though ratio of cross sections is still very insensitive
to the scale choice). Similar conclusion was reached in case of $\pi^0$ 
production in hadronic collisions \cite{se}. Comparison of PHENIX data for 
 inclusive $\pi^0$ 
production in hadronic collisions shows very good agreement with our NLO 
predictions obtained with all scales set to be equal to $p_T$ \cite{rhic200}.  
The ratio of $\pi^0$ production in heavy-ion collisions to the one in 
proton-proton collision is less sensitive to the change of scales, and 
in case of BH energy loss, the uncertainty is only a few percent.  
 
\begin{figure}[htp]
\centering
\setlength{\epsfxsize=7.5cm}
\centerline{\epsffile{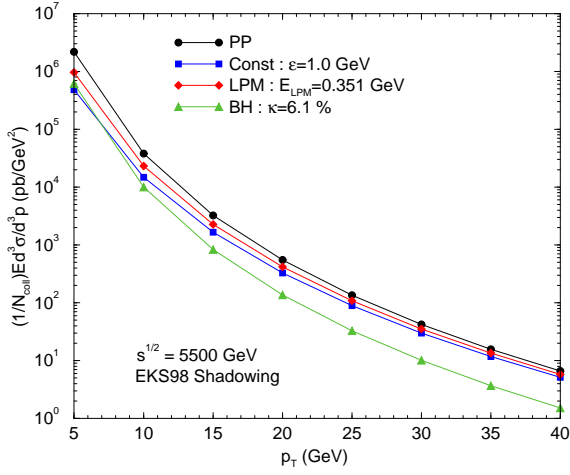}}
\caption{Inclusive $\pi^0$ cross sections at LHC energy, 
$\sqrt{s}=5.5$ TeV.} 
\label{fig:lhcdsigma}
\end{figure}

\begin{figure}[htp]
\centering
\setlength{\epsfxsize=7.5cm}
\centerline{\epsffile{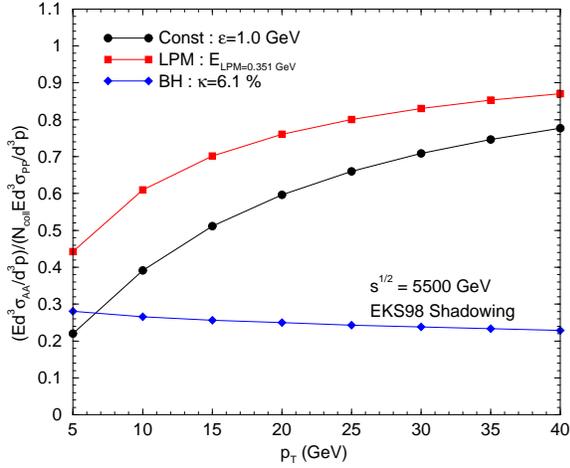}}
\caption{Ratio of inclusive $\pi^0$ cross sections in 
heavy ion and p-p collisions at LHC energy, $\sqrt{s}=5.5$ TeV.}
\label{fig:lhcratio}
\end{figure}

In Fig. (\ref{fig:lhcdsigma}) we show the inclusive $\pi^0$ cross sections
in p-p and heavy ion collisions at LHC energy of $\sqrt{s}=5.5$ TeV.
In case of heavy ion collisions, we show results again for three different
forms of energy loss parameter $\epsilon$ and include nuclear shadowing 
using the EKS98 parametrization. In Fig. (\ref{fig:lhcratio}), we present 
the ratio of $\pi^0$ inclusive cross section in heavy ion collisions to p-p.
Again, BH energy loss leads to a very different behavior of this ratio
as compared to constant and LPM energy loss. While this ratio has steeply 
increasing 
$p_T$ dependence in the case of constant and LPM energy loss, it is almost
flat in a large $p_T$ region in the case of BH energy loss. Again, nuclear
shadowing effects are small due to the large $p_T$'s involved. 

\begin{figure}[htp]
\centering
\setlength{\epsfxsize=7.5cm}
\centerline{\epsffile{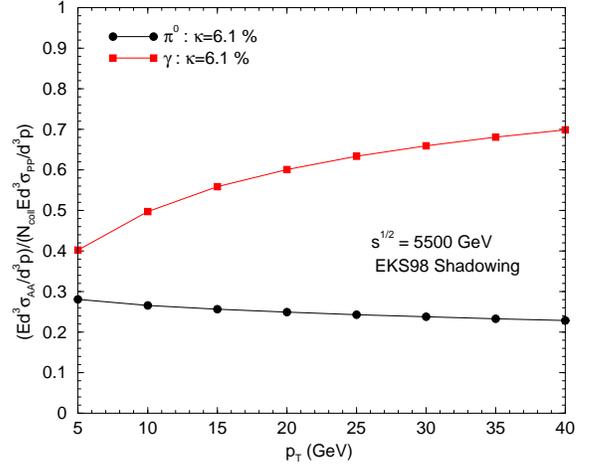}}
\caption{Prompt photon and inclusive $\pi^0$ cross sections in heavy
ion collisions normalized to p-p at $\sqrt{s}=5.5$ TeV.}
\label{fig:lhcgratio_pratio}
\end{figure}

In Fig. (\ref{fig:lhcgratio_pratio}) we show cross sections for prompt 
photon and inclusive $\pi^0$ production normalized to p-p using the BH
form of energy loss. Here we note that suppression of prompt photons is 
comparable to the $\pi^0$ case because at LHC energies, prompt photon 
production is dominated by bremsstrahlung processes (about $60\%$ 
contribution at $p_T \sim 4 GeV$) which are modified due to the energy 
loss in a similar way to the $\pi^0$ case. 

\begin{figure}[htp]
\centering
\setlength{\epsfxsize=7.5cm}
\centerline{\epsffile{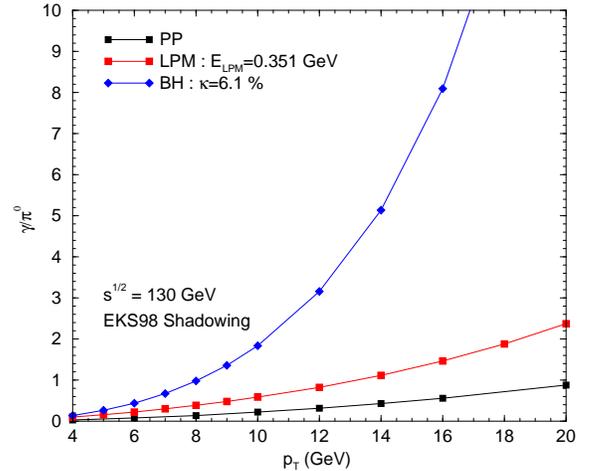}}
\caption{Ratio of prompt photon to $\pi^0$ cross sections at 
$\sqrt{s}=130$ GeV.}
\label{fig:rhic130_goverp}
\end{figure}

\begin{figure}[htp]
\centering
\setlength{\epsfxsize=7.5cm}
\centerline{\epsffile{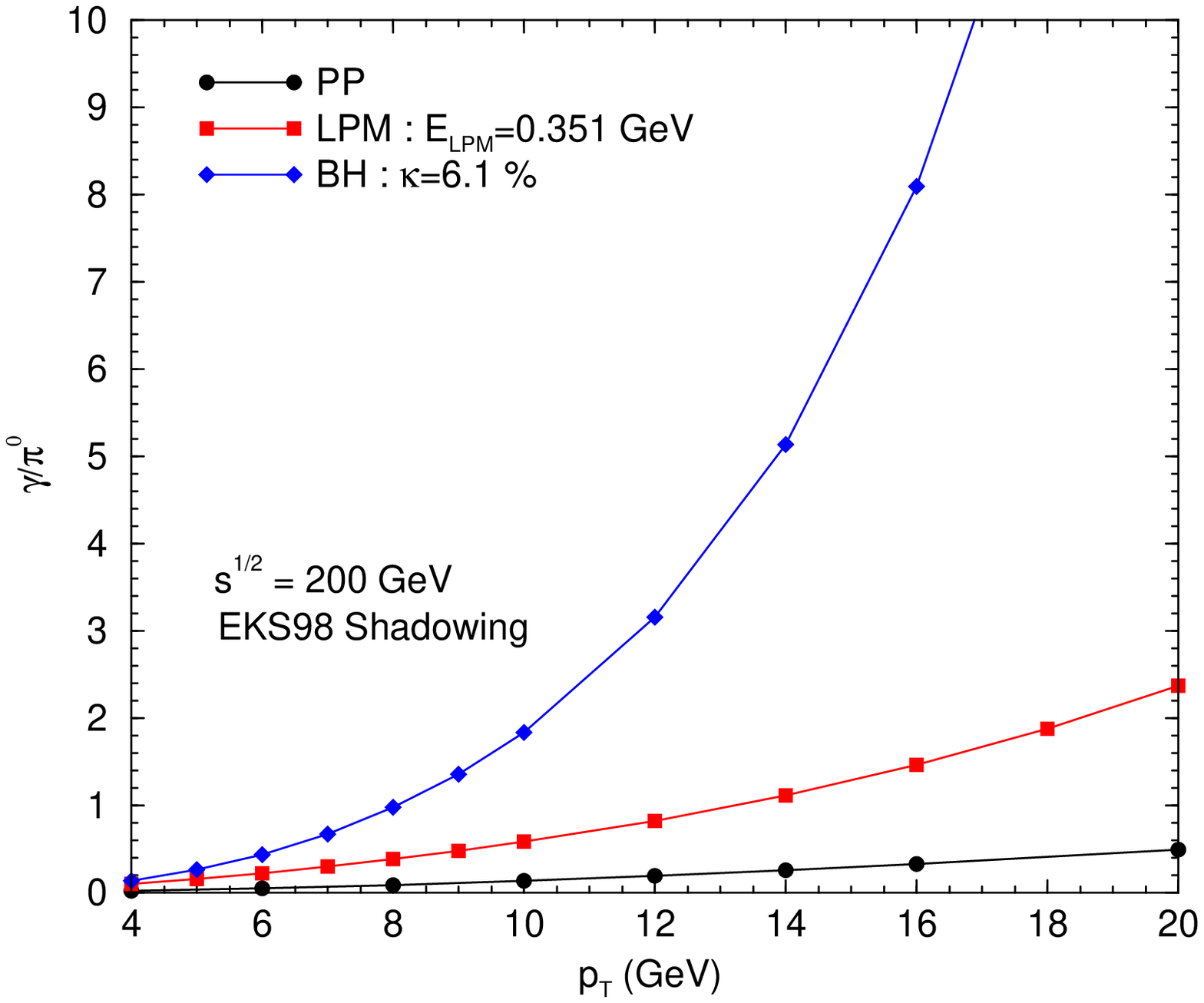}}
\caption{Ratio of prompt photon to $\pi^0$ cross sections at 
$\sqrt{s}=200$ GeV.}
\label{fig:rhic200_goverp}
\end{figure}

\begin{figure}[htp]
\centering
\setlength{\epsfxsize=7.5cm}
\centerline{\epsffile{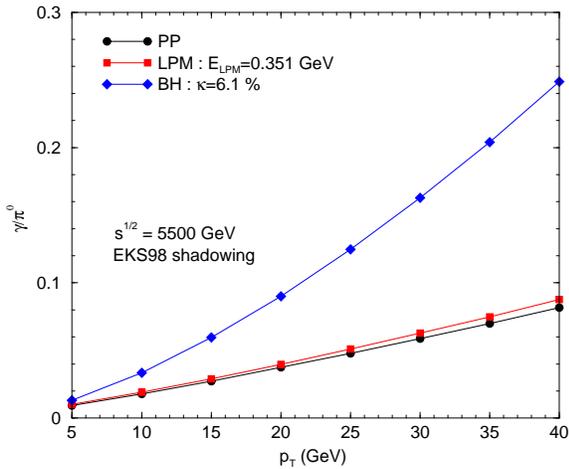}}
\caption{Ratio of prompt photon to $\pi^0$ cross sections at 
$\sqrt{s}=5.5$ TeV.}
\label{fig:lhc_goverp}
\end{figure}

Finally, in Figs. (\ref{fig:rhic130_goverp}, \ref{fig:rhic200_goverp}, 
\ref{fig:lhc_goverp})  we show the ratio of prompt photons to $\pi^0$ 
as a function of transverse momentum for energies $\sqrt{s}=130$ GeV, 
$\sqrt{s}=200$ GeV and $\sqrt{s}=5.5$ TeV, including nuclear shadowing and 
LPM as well as BH forms of energy loss.  We also show this ratio for 
proton-proton collisions at the same energy.  We find that in heavy-ion 
collisions, the $p_T$ 
dependence of the 
$\gamma/\pi^0$ ratio depends strongly on the form of the energy loss.  
For example, in case of LPM energy loss, this ratio increases with 
$p_T$ slowly, similar to the case in proton-proton collisions.  However, 
in case of BH energy loss, i.e. for 
$\epsilon=6.1\%E$ GeV, we find that 
because of the large $\pi^0$ 
suppression relative to prompt photons at RHIC energies, this 
ratio increases rapidly with $p_T$ approaching $1$ at $p_T\sim 8$ GeV.  
Similarly,  at the LHC 
this ratio increases more than order of magnitude, namely 
from 0.01 at low $p_T$ to about 0.25 at large 
$p_T$.  This is a much steeper increase than in case of proton-proton 
collisions, which increases only by a factor of 8 in the same $p_T$ range.  

\section{Summary and Conclusions}

We have calculated inclusive pion and prompt photon production
cross sections in heavy ion and hadronic collisions at RHIC
and LHC energies. We have incorrporated next-to-leading order 
contributions,  
initial state parton distribution functions in a nucleus and 
 medium induced parton energy loss by modifying the 
final state pion fragmentation function.  

We have shown that the nuclear K-factor, the ratio of NLO to LO
pion cross sections, is large and $p_T$ dependent. This $p_T$ dependence
is specially strong at lower energies, i.e. RHIC at $\sqrt{s} = 130$ GeV.
We find that nuclear shadowing effects are quite small, specially at
RHIC at all $p_T$'s considered. The EKS98 parametrization of nuclear
shadowing has a strong $p_T^2$ dependence which affects the cross sections 
through dependence of the factorization scale on $p_T$. If one could 
measure the $\pi^0$ spectra at very large $p_T$ at RHIC to high accuracy
(this would require an accuracy of better than $5\%$ at $p_t\sim 20$ GeV
which is quite unlikely), one could in principle rule out some models of 
nuclear shadowing which are $Q^2$ independent.  

We have investigated dependence of our results on three different forms
of energy loss. We considered constant energy loss as well as LPM and BH
type of energy loss. The medium induced energy loss effects in pion 
production are large even at $\sqrt{s} =130$ GeV. However, since our 
perturbative calculation is not applicable at low $p_T$, and the 
RHIC data at $\sqrt{s}=130 GeV$ is available for $p_T<4$GeV, 
 we can not distinguish
between different functional forms of energy loss from RHIC data at
$\sqrt{s}=130 GeV$. Therefore, we use the existing data at RHIC at 
$\sqrt{s} =200$ GeV to determine the functional form and magnitude of 
the average energy loss per collision, $\epsilon$.  We find that 
BH energy loss describes the data the best.  
We then predict the prompt photon 
production cross section in Au-Au collisions normalized to the p-p 
case at high $p_T$ ($4$ GeV$ <p_T<20$ GeV). We also predict 
the inclusive $\pi^0$ production cross section in heavy ion collisions 
normalized to p-p at the LHC energy, $\sqrt{s} =5.5$, assuming BH energy 
loss.  
  We demonstrate theoretical uncertainties
due to scale dependence of perturbative calculations of this ratio are
small ($<15\%$ between $Q=p_T/2$ and $Q=2p_T$).  

Finally, we show results for the ratio of prompt photon to $\pi^0$ cross 
sections at RHIC and LHC, of relevance to separating different sources of 
photon production. We show that contribution of direct photons is large 
at RHIC (about $75\%$ at $p_T=4 GeV$) but decreases as the energy is 
increasing, becoming $40\%$ at $p_T=4 GeV$ (with $60\%$ coming from 
bremsstrahlung processes) at LHC energies.

\vskip 0.1true in

\leftline{\bf Acknowledgments}

We would like to thank P. Aurenche and J. P. Guillet for providing us with
the fortran routines for calculating differential distributions for 
$\pi^0$ and photon 
production in hadronic collisions and for many useful discussions.  
We would like to thank D. d'Enterria and M. Tannenbaum for many helpful 
discussions and suggestions. We would also like to thank ITP, Santa Barbara 
where part of this work was done. J. J-M. would like to thank the
nuclear theory group at SUNY Stony Brook for their hospitality where
part of this work was done and is grateful to LBNL nuclear theory 
group for the use of their computing resources. This work was supported in 
part through U.S. Department of Energy Grants Nos. DE-FG03-93ER40792 and 
DE-FG02-95ER40906. S.J. is supported in part by the Natural Sciences and
 Engineering Research Council of Canada and by le Fonds pour la Formation
 de Chercheurs et l'Aide \`a la Recherche du Qu\'ebec.  
J.J-M. is supported in part by a PDF from BSA and by 
U.S. Department of Energy under Contract No. DE-AC02-98CH10886.

\end{document}